\documentclass{article}
\usepackage{graphicx}
\usepackage{amsmath}

\input{tcilatex}

\begin{document}

\title{Dilemma and Quantum Battle of Sexes}
\author{Ahmad Nawaz\thanks{%
Email address: el1anawaz@qau.edu.pk} and A.H. Toor\thanks{%
Email Address: ahtoor@qau.edu.pk} \\
Department of Electronics, \\
Quaid-i-Azam University, Islamabad 45320, Pakistan.}
\date{\today}
\maketitle

\begin{abstract}
We analyzed quantum version of the game Battle of Sexes using a general
initial quantum state. For a particular choice of initial entangled quantum
state it is shown that the classical dilemma of the Battle of Sexes can be
resolved and a unique solution of the game can be obtained.
\end{abstract}

\section{Introduction}

Game theory deals with situations where two or more players compete to
maximize their respective payoff or gain. Their interaction is strategic in
nature as their gain depends on the strategy adopted by other player/players 
\cite{von}. In the games with complete information, players decide their
strategy on the basis of payoff matrix, known to them. Nash Equilibrium (NE)
is a key concept, which is a set of strategies from which unilateral
deviation by a player reduces his/her payoff \cite{nash}.

In an important two player game, Battle of Sexes, Alice and Bob try to
decide a place to spend Saturday evening together. Alice wants to go for the
Opera and Bob is interested in watching TV while both prefer to spend their
evening together. The two NE of the game correspond to situations when both
the players choose either Opera or TV. Alice prefers NE based on Opera and
Bob prefers the other NE. In absence of any communication the two players
face a dilemma in choosing between the two NE and could inadvertently end up
with mismatch strategies. This mismatched strategy results in a loss for
both the players as they will not be able to spend the evening together and
termed as the worst payoff for both the players.

Extension of game theory to quantum domain \cite{meyer} leads to some
interesting new results in addition to the resolution of some existing
dilemmas in the classical versions of games. In quantum game theory an
initial quantum state is prepared by the arbiter and passed on to the
players. After applying their respective local operators (or strategies)
they return it to the arbiter who then determines their payoffs.$\emph{\ }$%
Initial quantum state plays a crucial role and interesting results are
obtained for initially entangled quantum states. Since players are equipped
with local operators they are unable to determine the complete initial
quantum state given to them and hence choose their strategies to maximize
their payoff on the basis of payoff matrix known to them. In an interesting
example, Eisert et. al. \cite{eisert} examined the game Prisoner Dilemma, in
quantum domain and showed that the dilemma which exists in the classical
version of the game does not exist in quantum version. Further they
constructed a quantum strategy which always wins over any classical
strategy. Inspired by their work, Marinatto and Weber \cite{marin} proposed
another interesting scheme to quantize the game of Battle of Sexes. They
introduced Hilbert structure to the strategic space of the game and argued
that if the players are allowed to play quantum strategies involving unitary
operators then the game has a unique solution, and the dilemma could be
resolved. They used maximally entangled initial quantum state, and allowed
the players to play strategies which are combination of the identity
operator $I$ and the flip operator $C$, with classical probabilities $%
p^{\ast }$ and $\left( 1-p^{\ast }\right) $ for Alice and $q^{\ast }$ and $%
\left( 1-q^{\ast }\right) $ for Bob. In the quantum version of the game
maximum payoff is obtained for the two pure\emph{\ }NE,\textbf{\ }$\mathbf{(}%
p^{\ast }=q^{\ast }=1)$ or $\left( 1,1\right) $ and $\mathbf{(}0,0)$. In
Marinatto and Weber scheme both NE corresponds to equal payoff and hence
they argued that the classical dilemma is no longer there as both the
solutions or NE are equally good for the two players.

In an interesting comment Benjamin \cite{benjamin} pointed out that the
dilemma still exists as the same payoff for the two NE make them equally
acceptable to both the players and there is no way for them to prefer ''$1$%
'' over ''$0$'' , in absence of any communication between them. He argued
that the players still face some what similar dilemma as they could still
end up with a situation $\left( 1,0\right) $ or $\left( 0,1\right) $ which
corresponds to the worst payoff for the both. In their response to
Benjamin's comment, Marinatto and Weber \cite{marin1} insisted that since
both the NE $\left( 0,0\right) $ and $\left( 1,1\right) $ render the initial
quantum state unchanged and corresponds to equal and maximum payoff for both
the players, therefore, both the players would prefer $\left( 1,1\right) ,$
as by choosing $p$ or $q$ equal to zero there is a danger for both the
payers to get in to a situation $\left( 1,0\right) $ or $\left( 0,1\right) $
which corresponds to the lowest payoff. However, choosing strategy on this
argument requires complete information of the initial quantum state and in
quantum games players can not measure the initial quantum state\cite{azhar} 
\cite{lee}\cite{witte}. \emph{\ }

In this paper we consider a general initial quantum state and present a
condition on the parameters of the initial quantum state for which classical
dilemma can be resolved and a unique solution of the quantum Battle of Sexes
can be obtained. In comparison to Marinatto and Weber \cite{marin} we
presented a condition for which payoffs corresponding to ``mismatched or
worst case situation'' are different for the two players which results in a
unique solution of the game. Classical and quantum versions of the game of
Battle of Sexes are presented in Section 2. Here we restrict our analysis to
pure strategies only as classical dilemma deals with these strategies only.

\section{Battle of Sexes}

\subsection{\label{classical} \textbf{Classical Form}}

Battle of Sexes is an interesting static game. In the usual exposition of
this game two players Alice and Bob try to decide a place to spend Saturday
evening together. Alice wants to go for the Opera while Bob is interested in
watching TV and both would like to be together. The game is represented by
the following payoff matrix 
\begin{equation}
\text{Alice} 
\begin{array}{c}
O \\ 
T
\end{array}
\overset{\text{Bob}}{\overset{
\begin{array}{cc}
O\text{ \ \ \ \ } & T
\end{array}
}{\left[ 
\begin{array}{cc}
\left( \alpha ,\beta \right) & \left( \gamma ,\gamma \right) \\ 
\left( \gamma ,\gamma \right) & \left( \beta ,\alpha \right)
\end{array}
\right] }},  \label{classical payoff}
\end{equation}
where $O$ and $T$ represent Opera and TV, respectively. The elements $\alpha 
$, $\beta $, $\gamma $ are the payoffs for the players corresponding to the
choices available to them with $\alpha >\beta >\gamma $. The two pure\emph{\ 
}NE of this game are $(O,O)$ and $(T,T)$ which corresponds to the situation
when both the players choose Opera and TV, respectively. Here the first NE
is more favorable to Alice while the second NE is favorable to Bob. Since
the players are not allowed to communicate, they face a dilemma in choosing
their strategy. The strategy pairs $\left( O,T\right) $ and $\left(
T,O\right) $ correspond to the worst-case payoff for the two players, i.e.,
both the players gets\emph{\ }the minimum possible\emph{\ }payoff $\gamma .$
\ There also exists a mixed NE for this game but we are not interested in it
here.

\subsection{\label{quantum} \textbf{Quantum Form}}

In the quantum version of the game both players, Alice and Bob, apply their
respective strategy on the initial quantum state given to them on the basis
of payoff matrix known to them. In this approach the payoff matrix depends
on the initial state and can be controlled by the parameters of the initial
quantum state. Our choice of general initial quantum state provide us with
additional parameters to control the game in comparison to Marinatto and
Webers \cite{marin}.

Let Alice and Bob have the following initial entangled quantum state at
their disposal:

\begin{equation}
\left| \psi _{in}\right\rangle =a\left| OO\right\rangle +b\left|
OT\right\rangle +c\left| TO\right\rangle +d\left| TT\right\rangle ,
\label{fun}
\end{equation}
where \ $\left| a\right| ^{2}+\left| b\right| ^{2}+\left| c\right|
^{2}+\left| d\right| ^{2}=1.$ Here the first entry in ket $\left|
{}\right\rangle $ is for Alice and the second for Bob's strategy. For $b$
and $c$ equal to zero, the eq. (\ref{fun}) reduces to the initial entangled
quantum state used by Marinatto and Weber \cite{marin}. The unitary
operators at the disposal of the two players are defined as:

\begin{eqnarray}
C\left| O\right\rangle &=&\left| T\right\rangle ,\text{ \ \ }C\left|
T\right\rangle =\left| O\right\rangle ,\text{ \ \ }C=C^{\dagger }=C^{-1}, 
\notag \\
I\left| O\right\rangle &=&\left| O\right\rangle ,\text{ \ \ }I\left|
T\right\rangle =\left| I\right\rangle ,\text{ \ \ \ \ }I=I^{\dagger }=I^{-1}.
\label{oper}
\end{eqnarray}
Following the Marinatto and Weber's approach, take $pI+\left( 1-p\right) C$
and $qI+\left( 1-q\right) C$ as the strategies for the two players,
respectively, with $p$ and $q$ being the classical probabilities for using
the identity operator $I.$ The final density matrix takes the form: 
\begin{gather}
\rho _{f}=pqI_{A}\otimes I_{B}\rho _{in}I_{A}^{\dagger }\otimes
I_{B}^{\dagger }+p(1-q)I_{A}\otimes C_{B}\rho _{in}I_{A}^{\dagger }\otimes
C_{B}^{\dagger }  \notag \\
+q(1-p)C_{A}\otimes I_{B}\rho _{in}C_{A}^{\dagger }\otimes I_{B}^{\dagger
}+(1-p)(1-q)C_{A}\otimes C_{B}\rho _{in}C_{A}^{\dagger }\otimes
C_{B}^{\dagger }.  \label{def}
\end{gather}
Here $\rho _{in}=\left| \psi _{in}\right\rangle \left\langle \psi
_{in}\right| $ which can be obtained through the eq. (\ref{fun}) (See
Appendix A)$.$ The corresponding payoff operators for Alice and Bob are:

\begin{eqnarray}
P_{A} &=&\alpha \left| OO\right\rangle \left\langle OO\right| +\beta \left|
TT\right\rangle \left\langle TT\right| +\gamma (\left| OT\right\rangle
\left\langle OT\right| +\left| TO\right\rangle \left\langle TO\right| ),
\label{popa} \\
P_{B} &=&\beta \left| OO\right\rangle \left\langle OO\right| +\alpha \left|
TT\right\rangle \left\langle TT\right| +\gamma (\left| OT\right\rangle
\left\langle OT\right| +\left| TO\right\rangle \left\langle TO\right| ),
\label{popb}
\end{eqnarray}
and the payoff functions (the mean values of these operators, i.e., $%
\$_{A}(p,q)=$ Trace$[P_{A}\rho _{f}]$ and $\$_{B}(p,q)=$Trace$[P_{B}\rho
_{f}]$) are obtained by using eqs. (\ref{oper},\ref{def},\ref{popa},\ref
{popb}) and are given as 
\begin{gather}
\$_{A}(p,q)=p\left[ q\Omega +\Phi \left( \left| b\right| ^{2}-\left|
d\right| ^{2}\right) +\Lambda \left( \left| c\right| ^{2}-\left| a\right|
^{2}\right) \right]  \notag \\
+q\left[ \Lambda \left( \left| b\right| ^{2}-\left| a\right| ^{2}\right)
+\Phi \left( \left| c\right| ^{2}-\left| d\right| ^{2}\right) \right]
+\Theta ,  \label{payoff1}
\end{gather}
\begin{gather}
\$_{B}(p,q)=q\left[ p\Omega +\Phi \left( \left| b\right| ^{2}-\left|
a\right| ^{2}\right) +\Lambda \left( \left| c\right| ^{2}-\left| d\right|
^{2}\right) \right]  \notag \\
+p\left[ \Lambda \left( \left| b\right| ^{2}-\left| d\right| ^{2}\right)
+\Phi \left( \left| c\right| ^{2}-\left| a\right| ^{2}\right) \right]
+\Theta .  \label{payoff2}
\end{gather}
In writing above equations we have used defined 
\begin{eqnarray}
\QTR{sl}{\ }\Omega &=&(\alpha +\beta -2\gamma )(\left| a\right| ^{2}-\left|
b\right| ^{2}-\left| c\right| ^{2}+\left| d\right| ^{2}),  \notag \\
\Phi &=&(\alpha -\gamma )\QTR{sl}{,\ }\Lambda =(\beta -\gamma ),  \notag \\
\Theta &=&\alpha \left| d\right| ^{2}+\gamma \left| c\right| ^{2}+\gamma
\left| b\right| ^{2}+\beta \left| a\right| ^{2}.  \label{define}
\end{eqnarray}
The NE of the game are found by solving the following two inequalities: 
\begin{eqnarray}
\$_{A}(p^{\ast },q^{\ast })-\$_{A}(p,q^{\ast }) &\geq &0,  \notag \\
\$_{B}(p^{\ast },q^{\ast })-\$_{B}(p,q^{\ast }) &\geq &0,
\end{eqnarray}
that lead to the following two conditions, respectively: 
\begin{gather}
(p^{\ast }-p)[q^{\ast }(\alpha +\beta -2\gamma )(\left| a\right| ^{2}-\left|
b\right| ^{2}-\left| c\right| ^{2}+\left| d\right| ^{2})+  \notag \\
(\gamma -\beta )\left| a\right| ^{2}+(\alpha -\gamma )\left| b\right|
^{2}+(\beta -\gamma )\left| c\right| ^{2}+(\gamma -\alpha )\left| d\right|
^{2}]\geq 0,  \label{nash1}
\end{gather}
and

\begin{gather}
(q^{\ast }-q)[p^{\ast }(\alpha +\beta -2\gamma )(\left| a\right| ^{2}-\left|
b\right| ^{2}-\left| c\right| ^{2}+\left| d\right| ^{2})+  \notag \\
(\gamma -\alpha )\left| a\right| ^{2}+(\alpha -\gamma )\left| b\right|
^{2}+(\beta -\gamma )\left| c\right| ^{2}+(\gamma -\beta )\left| d\right|
^{2}]\geq 0.  \label{nash2}
\end{gather}
The above two inequalities are satisfied if both the factors have the same
sign. Here we are interested in solving the dilemma arising due to pure
strategies, i.e. $\left( 1,1\right) $ and $\left( 0,0\right) $, therefore,
we restrict ourselves to the following possible pure strategies pairs:

\textbf{Case (a) }When $p^{\ast }=0,q^{\ast }=0$ then the inequalities (\ref
{nash1}) and (\ref{nash2}), reduce to

\begin{eqnarray}
(\gamma -\beta )\left| a\right| ^{2}+(\alpha -\gamma )\left| b\right|
^{2}+(\beta -\gamma )\left| c\right| ^{2}+(\gamma -\alpha )\left| d\right|
^{2} &<&0,  \notag \\
(\gamma -\alpha )\left| a\right| ^{2}+(\alpha -\gamma )\left| b\right|
^{2}+(\beta -\gamma )\left| c\right| ^{2}+(\gamma -\beta )\left| d\right|
^{2}] &<&0.  \label{cond1}
\end{eqnarray}
All those values of the initial quantum state parameters for which above
inequalities are satisfied, strategy pair $\left( 0,0\right) $ is a Nash
equilibrium. Here we consider a particular set of values for the initial
state parameters for which a unique solution of the game can be found and
hence dilemma would be resolved, however, this choice is not unique. Lets
take

\begin{equation}
\left| a\right| ^{2}=\left| d\right| ^{2}=\left| b\right| ^{2}=\frac{5}{16}%
,\left| c\right| ^{2}=\frac{1}{16}.  \label{para}
\end{equation}
Corresponding payoffs can be obtained from Eqs. (\ref{payoff1}) and (\ref
{payoff2}) are

\begin{eqnarray}
\$_{A}(0,0) &=&\frac{5\alpha +5\beta +6\gamma }{16},  \notag \\
\$_{B}(0,0) &=&\frac{5\alpha +5\beta +6\gamma }{16}.  \label{zero}
\end{eqnarray}
Physically it means that for the NE $\left( 0,0\right) ,$ two players get
equal payoff corresponding to the choice of initial state parameters given
by the eq.(\ref{para}).

\textbf{Case (b): }When $p^{\ast }=q^{\ast }=1,$ then the inequalities (\ref
{nash1}) and (\ref{nash2}), become,

\begin{eqnarray}
(\alpha -\gamma )\left| a\right| ^{2}+(\gamma -\beta )\left| b\right|
^{2}+(\gamma -\alpha )\left| c\right| ^{2}+(\beta -\gamma )\left| d\right|
^{2} &>&0,  \notag \\
(\beta -\gamma )\left| a\right| ^{2}+(\gamma -\beta )\left| b\right|
^{2}+(\gamma -\alpha )\left| c\right| ^{2}+(\alpha -\gamma )\left| d\right|
^{2} &>&0.  \label{cond2}
\end{eqnarray}
These inequalities are again satisfied for the choice of parameters given in
the eq. (\ref{para}) for the initial quantum state and the strategy pair $%
\left( 1,1\right) $ is also a NE.\emph{\ }Corresponding payoffs for the two
players in this case are: 
\begin{eqnarray}
\$_{A}(1,1) &=&\frac{5\alpha +5\beta +6\gamma }{16},  \notag \\
\$_{B}(1,1) &=&\frac{5\alpha +5\beta +6\gamma }{16}.  \label{first}
\end{eqnarray}
For the mismatched strategy pairs, i.e., $(p^{\ast }=0,q^{\ast }=1)$ and $%
(p^{\ast }=1,q^{\ast }=0)$ the inequalities (\ref{nash1}) and (\ref{nash2})
are not satisfied for the choice of initial state parameters given by the
eq. (\ref{para}), hence these strategy pairs are not NE. However, it is
interesting to note the corresponding payoffs for the two players, i.e.,

\begin{eqnarray}
\$_{A}(0,1) &=&\frac{\alpha +5\beta +10\gamma }{16},\text{\ \ \ }\$_{B}(0,1)=%
\frac{5\alpha +\beta +10\gamma }{16},  \notag \\
\$_{A}(1,0) &=&\frac{5\alpha +\beta +10\gamma }{16},\text{ \ \ }\$_{B}(1,0)=%
\frac{\alpha +5\beta +10\gamma }{16}.  \label{worst-1}
\end{eqnarray}
Now keeping in view all the payoffs given by Eqs. (\ref{zero},\ref{first},%
\ref{worst-1}), under the choice (\ref{para}), the quantum game can be
represented by the following payoff matrix:

\begin{equation}
\text{Alice} 
\begin{array}{c}
p=1 \\ 
p=0
\end{array}
\overset{\text{Bob}}{\overset{
\begin{array}{cc}
q=1\text{ \ \ \ \ } & q=0
\end{array}
}{\left[ 
\begin{array}{cc}
\left( \alpha ^{\prime },\alpha ^{\prime }\right) & \left( \beta \prime
,\gamma ^{\prime }\right) \\ 
\left( \gamma ^{\prime },\beta ^{\prime }\right) & \left( \alpha ^{\prime
},\alpha ^{\prime }\right)
\end{array}
\right] }},  \label{matrix1}
\end{equation}
where 
\begin{eqnarray}
\alpha ^{\prime } &=&\frac{5\alpha +5\beta +6\gamma }{16},  \notag \\
\beta ^{\prime } &=&\frac{5\alpha +\beta +10\gamma }{16},  \notag \\
\gamma ^{\prime } &=&\frac{\alpha +5\beta +10\gamma }{16},
\end{eqnarray}
with $\alpha ^{\prime }>\beta ^{\prime }>\gamma ^{\prime }.$ On the other
hand, quantized version of Marinatto and Weber \cite{marin} can be
represented by the following payoff matrix:

\begin{equation}
\text{Alice}\ \ \ 
\begin{array}{c}
p=1 \\ 
p=0
\end{array}
\overset{\text{Bob}}{\overset{
\begin{array}{cc}
q=1\text{ \ \ \ \ } & q=0
\end{array}
}{\left[ 
\begin{array}{cc}
\left( \frac{\alpha +\beta }{2},\frac{\alpha +\beta }{2}\right) & \left(
\gamma ,\gamma \right) \\ 
\left( \gamma ,\gamma \right) & (\frac{\alpha +\beta }{2},\frac{\alpha
+\beta }{2})
\end{array}
\right] .}}  \label{matrix2}
\end{equation}
In comparison to classical version payoff matrix, i.e., eq. (\ref{classical
payoff}), both Marinatto and Weber's payoff matrix (\ref{matrix2}) and our
payoff matrix (\ref{matrix1}) shows a clear advantage over the classical
version as the payoffs for the two players are same for the two pure NE in
quantum versions of the game. Hence there is no incentive for the two
players to prefer one NE over the other. However, as pointed out by Benjamin 
\cite{benjamin}, in the Marinatto's quantum version, in absence of any
communication the two players could inadvertently end up with a mismatched
strategies, i.e., $\left( 1,0\right) $ or $\left( 0,1\right) $ which
corresponds to minimum possible payoff, i.e., $\gamma ,$ for both the
players. It is important to note that in our version of the quantum Battle
of Sexes the payoffs corresponding to worst-case situation are different for
the two players. This particular feature leads to a unique solution for the
game by providing a straight forward reason for rational players to go for
one of the NE, i.e., $\left( 1,1\right) $ for the particular choice of
parameters given by the eq. (\ref{para}).

It can be seen from the payoff matrix (\ref{matrix1}), that the payoff for
the two players is maximum for the two NE, $\left( 0,0\right) $ and $\left(
1,1\right) $, but for Alice the rational choice is $p^{\ast }=1$ since her
payoff is maximum, i.e., $\alpha ^{\prime },$ when Bob decides to play $%
q^{\ast }=1$ and equals to $\beta ^{\prime }$ if Bob decides to play $%
q^{\ast }=0,$ which is higher than the worst possible payoff, i.e., $\gamma
^{\prime }$. In a similar manner for Bob the rational choice is $q^{\ast }=1$
since his payoff is maximum, i.e., $\alpha ^{\prime },$ when Alice also
plays $p^{\prime }=1$ and equals to $\beta ^{\prime }$ when Alice plays $%
p^{\prime }=0$ which better than the worst possible. Thus for the initial
quantum with parameters given by the eq.(\ref{para}), NE $\left( 1,1\right) $
is clearly a preferred strategy for both the players giving a unique
solution to the game. Similarly an initial quantum state, for example, with
parameters $\left| a\right| ^{2}=\left| d\right| ^{2}=\left| c\right| ^{2}=%
\frac{5}{16},$ $\left| b\right| ^{2}=\frac{1}{16}$\ can be found for which $%
\left( 0,0\right) $ is a preferred strategy pair for both the players giving
a unique solution for the game.

\section{Summary}

We analyzed the game of quantum Battle of Sexes using the approach developed
by Marinatto and Weber \cite{marin}. Instead of restricting to the maximally
entangled initial quantum state we considered a general initial quantum
state. Exploiting the additional parameters in the initial state we
presented a condition for which unique solution of the game can be obtained.
In particular we addressed the issues pointed out by Benjamin \cite{benjamin}
in Marinatto and Weber \cite{marin} quantum version of the Battle of sexes
game. In our approach difference in the payoffs for the two players
corresponding to so called worst-case situation leads to a unique solution
of the game. Our results reduces to that of Marinatto and Weber under
appropriate conditions.

\appendix

\section{Appendix}

\subsubsection{Derivation of Initial Density Matrix}

The initial quantum state, i.e. eq. (\ref{fun}), can be used to obtain the
required initial density matrix

\begin{equation}
\rho _{in}=\left( a\left| OO\right\rangle +b\left| OT\right\rangle +c\left|
TO\right\rangle +d\left| TT\right\rangle \right) \left( a^{\ast
}\left\langle OO\right| +b^{\ast }\left\langle TO\right| +c^{\ast
}\left\langle OT\right| +d^{\ast }\left\langle TT\right| \right)
\label{ini-density}
\end{equation}

where $\ast $ stands for complex conjugate. Density matrix can also be
written in the following form: 
\begin{gather}
\rho _{in}=\left| a\right| ^{2}\left| OO\right\rangle \left\langle OO\right|
+ab^{\ast }\left| OO\right\rangle \left\langle TO\right| +ac^{\ast }\left|
OO\right\rangle \left\langle OT\right| +ad^{\ast }\left| OO\right\rangle
\left\langle TT\right|  \notag \\
+ba^{\ast }\left| OT\right\rangle \left\langle OO\right| +\left| b\right|
^{2}\left| OT\right\rangle \left\langle TO\right| +bc^{\ast }\left|
OT\right\rangle \left\langle OT\right| +bd^{\ast }\left| OT\right\rangle
\left\langle TT\right|  \notag \\
+ca^{\ast }\left| TO\right\rangle \left\langle OO\right| +cb^{\ast }\left|
TO\right\rangle \left\langle TO\right| +\left| c\right| ^{2}\left|
TO\right\rangle \left\langle OT\right| +cd^{\ast }\left| TO\right\rangle
\left\langle TT\right|  \notag \\
+da^{\ast }\left| TT\right\rangle \left\langle OO\right| +db^{\ast }\left|
TT\right\rangle \left\langle TO\right| +dc^{\ast }\left| TT\right\rangle
\left\langle OT\right| +\left| d\right| ^{2}\left| TT\right\rangle
\left\langle TT\right|  \label{in-density}
\end{gather}

\subsubsection{Calculation for the Payoffs}

>From eq (\ref{def}) the final density operator is

\begin{gather}
\rho _{f}=pqI_{A}\otimes I_{B}\rho _{in}I_{A}^{\dagger }\otimes
I_{B}^{\dagger }+p(1-q)I_{A}\otimes C_{B}\rho _{in}I_{A}^{\dagger }\otimes
C_{B}^{\dagger }  \notag \\
+q(1-p)C_{A}\otimes I_{B}\rho _{in}C_{A}^{\dagger }\otimes I_{B}^{\dagger
}+(1-p)(1-q)C_{A}\otimes C_{B}\rho _{in}C_{A}^{\dagger }\otimes
C_{B}^{\dagger }.  \label{final-density}
\end{gather}
With the help of eq (\ref{oper}) $\rho _{f}$ becomes

\QTP{Body Math}
\begin{equation*}
\left[ 
\begin{array}{cccc}
\begin{array}{l}
pq\zeta +\left| d\right| ^{2} \\ 
+p(\left| b\right| ^{2}-\left| d\right| ^{2}) \\ 
+q(\left| c\right| ^{2}-\left| d\right| ^{2})
\end{array}
& 
\begin{array}{l}
pq\epsilon +dc^{\ast } \\ 
+p(ba^{\ast }-dc^{\ast }) \\ 
+q(cd^{\ast }-dc^{\ast })
\end{array}
& 
\begin{array}{l}
pq\omega +db^{\ast } \\ 
+p(ba^{\ast }-db^{\ast }) \\ 
+q(ca^{\ast }-db^{\ast })
\end{array}
& 
\begin{array}{l}
pq\xi +da^{\ast } \\ 
+p(cb^{\ast }-da^{\ast }) \\ 
+q(cb^{\ast }-da^{\ast })
\end{array}
\\ 
\begin{array}{l}
-pq\epsilon +cd^{\ast } \\ 
+p(ab^{\ast }-cd^{\ast }) \\ 
+q(dc^{\ast }-cd^{\ast })
\end{array}
& 
\begin{array}{l}
-pq\zeta +\left| c\right| ^{2} \\ 
+p(\left| a\right| ^{2}-\left| c\right| ^{2}) \\ 
+q(\left| d\right| ^{2}-\left| c\right| ^{2})
\end{array}
& 
\begin{array}{l}
-pq\xi +cb^{\ast } \\ 
+p(ad^{\ast }-cb^{\ast }) \\ 
+q(da^{\ast }-cb^{\ast })
\end{array}
& 
\begin{array}{l}
-pq\omega +ca^{\ast } \\ 
+p(ac^{\ast }-ca^{\ast }) \\ 
+q(db^{\ast }-ca^{\ast })
\end{array}
\\ 
\begin{array}{l}
-pq\omega +bd^{\ast } \\ 
+p(db^{\ast }-bd^{\ast }) \\ 
+q\left( ac^{\ast }-bd^{\ast }\right)
\end{array}
& 
\begin{array}{l}
-pq\xi +bc^{\ast } \\ 
+p(da^{\ast }-bc^{\ast }) \\ 
+q(ad^{\ast }-bc^{\ast })
\end{array}
& 
\begin{array}{l}
-pq\zeta +\left| b\right| ^{2} \\ 
+p(\left| d\right| ^{2}-\left| b\right| ^{2}) \\ 
+q(\left| a\right| ^{2}-\left| b\right| ^{2})
\end{array}
& 
\begin{array}{l}
-pq\epsilon +ba^{\ast } \\ 
+p(dc^{\ast }-ba^{\ast }) \\ 
+q(ab^{\ast }-ba^{\ast })
\end{array}
\\ 
\begin{array}{l}
pq\xi +ad^{\ast } \\ 
+p(cb^{\ast }-ad^{\ast }) \\ 
+q(bc^{\ast }-ad^{\ast })
\end{array}
& 
\begin{array}{l}
pq\omega +ac^{\ast } \\ 
+p(ca^{\ast }-ac^{\ast }) \\ 
+q(bd^{\ast }-ac^{\ast })
\end{array}
& 
\begin{array}{l}
pq\epsilon +ab^{\ast } \\ 
+p(cd^{\ast }-ab^{\ast }) \\ 
+q(ba^{\ast }-ab^{\ast })
\end{array}
& 
\begin{array}{l}
pq\zeta +\left| a\right| ^{2} \\ 
+p(\left| c\right| ^{2}-\left| a\right| ^{2}) \\ 
+q(\left| b\right| ^{2}-\left| a\right| ^{2})
\end{array}
\end{array}
\right]
\end{equation*}

where we defined

\begin{eqnarray*}
\epsilon &=&ab^{\ast }-ba^{\ast }-cd^{\ast }+dc^{\ast } \\
\zeta &=&\left| a\right| ^{2}-\left| b\right| ^{2}-\left| c\right|
^{2}+\left| d\right| ^{2} \\
\omega &=&ac^{\ast }-bd^{\ast }-ca^{\ast }+db^{\ast } \\
\xi &=&ad^{\ast }-bc^{\ast }-cb^{\ast }+da^{\ast }
\end{eqnarray*}
and the payoff functions (the mean values of these operators, i.e., $%
\$_{A}(p,q)=$ Trace$[P_{A}\rho _{f}]$ and $\$_{B}(p,q)=$Trace$[P_{B}\rho
_{f}]$) using eqs. (\ref{final-density}, \ref{popa},\ref{popb}) become:

\begin{eqnarray}
\$_{A}(p,q) &=&p[q(\alpha +\beta -2\gamma )(\left| a\right| ^{2}-\left|
b\right| ^{2}-\left| c\right| ^{2}+\left| d\right| ^{2})  \notag \\
&&+(\gamma -\beta )\left| a\right| ^{2}+(\alpha -\gamma )\left| b\right|
^{2}+(\beta -\gamma )\left| c\right| ^{2}+(\gamma -\alpha )\left| d\right|
^{2}]  \notag \\
&&+q[(\gamma -\beta )\left| a\right| ^{2}+(\beta -\gamma )\left| b\right|
^{2}+(\alpha -\gamma )\left| c\right| ^{2}  \notag \\
&&+(\gamma -\alpha )\left| d\right| ^{2}]+\alpha \left| d\right| ^{2}+\gamma
\left| c\right| ^{2}+\gamma \left| b\right| ^{2}+\beta \left| a\right| ^{2}
\label{a}
\end{eqnarray}

\begin{eqnarray}
\$_{B}(p,q) &=&q[p(\alpha +\beta -2\gamma )(\left| a\right| ^{2}-\left|
b\right| ^{2}-\left| c\right| ^{2}+\left| d\right| ^{2})  \notag \\
&&+(\gamma -\alpha )\left| a\right| ^{2}+(\alpha -\gamma )\left| b\right|
^{2}+(\beta -\gamma )\left| c\right| ^{2}+(\gamma -\beta )\left| d\right|
^{2}]  \notag \\
&&+p[(\gamma -\alpha )\left| a\right| ^{2}+(\beta -\gamma )\left| b\right|
^{2}+(\alpha -\gamma )\left| c\right| ^{2}  \notag \\
&&+(\gamma -\beta )\left| d\right| ^{2}]+\beta \left| d\right| ^{2}+\gamma
\left| c\right| ^{2}+\gamma \left| b\right| ^{2}+\alpha \left| a\right| ^{2}
\label{b}
\end{eqnarray}
\textsl{\ }The above equations leads to the eqs. (\ref{payoff1},\ref{payoff2}%
) in terms of quantities defined by eqs. (\ref{define}).

\end{document}